\documentclass[a4paper, 9pt, conference]{ieeeconf}      

\IEEEoverridecommandlockouts                              

\usepackage{graphics} 
\usepackage{epsfig} 
\usepackage{mathptmx} 
\usepackage{times} 
\usepackage{amsmath} 
\usepackage{amssymb}  
\usepackage{graphicx}
\usepackage{subfig}
\usepackage{algorithm}
\usepackage{algorithmic}
\usepackage{booktabs} 
\usepackage{pifont} 

\newcommand{\cmark}{\ding{51}} 
\newcommand{\xmark}{\ding{55}} 

\title{\LARGE \bf
 Advancements in Tactile Hand Gesture Recognition for Enhanced Human-Machine Interaction }

\author{Chiara Fumelli, Anirvan Dutta, and Mohsen Kaboli 
\thanks{All authors are with BMW Group Research, RoboTac Lab. e-mail: name.surname@bmwgroup.com.}
\thanks{Funded in part by the EU H2020 INTUITIVE under Grant ID 861166 and in part by EU Horizon PHASTRAC under Grant ID 101092096.
}
}

\begin{document}
\maketitle
\thispagestyle{empty}
\pagestyle{empty}

\begin{abstract}
Motivated by the growing interest in enhancing intuitive physical Human-Machine  Interaction (HRI/HVI), this study aims to propose a robust tactile hand gesture recognition system. We performed a comprehensive evaluation of different hand gesture recognition approaches for a large area tactile sensing interface (touch interface) constructed from conductive textiles.  Our evaluation encompassed traditional feature engineering methods, as well as contemporary deep learning techniques capable of real-time interpretation of a range of hand gestures, accommodating variations in hand sizes, movement velocities, applied pressure levels, and interaction points. Our extensive analysis of the various methods makes a significant contribution to tactile-based gesture recognition in the field of human-machine interaction.

%

\end{abstract}

\section{Introduction}
\label{intro}

The sense of touch plays a pivotal role in human interactions, enabling individuals to perceive and respond to the external environment \cite{kaboli2017tactile, kaboli_texture_tro, kaboli2019auro}. Interest in the study of human-robot interactions (HRI) and human-vehicle interactions (HVI) is growing, with a focus on creating more user-friendly, natural, and intuitive engagements \cite{murali2022intelligent}. This burgeoning interest propels research endeavours towards emulating human-like touch perception in machines, seeking to bridge the divide between human-human and human-machine interactions. 

Touch interfaces stand as primary mediums of interaction between humans and machines \cite{RecentProgressofTactile}, in which individuals engage through gestures using their fingers or hands. However, conventional methods and hardware are limited in their ability to recognize a wide range of gestures \cite{debard2018learning}. Furthermore, the touch interfaces commonly employed today exhibit constraints concerning dimensions, processing speed, and other factors. Efforts to cultivate a biomimetic sense of touch in recent decades have stimulated the exploration and development of diverse sensor technologies characterized by flexibility and scalability \cite{sandamirskaya2022neuromorphic}. One of the emerging touch technologies inspired by human skin is textile tactile interfaces \cite{TextileBased}, which can be deployed over large areas, intricate shapes, and with varying resolutions \cite{largeAreaSensor}. However, textile-based technologies encounter challenges such as efficient data and signal processing, deformation, high signal-to-noise ratio (SNR), hysteresis, etc. Despite substantial progress in this field, the integration of these interfaces into consumer electronics remains subpar compared to their conventional counterparts, which predominantly employ low-footprint rigid materials.

In this study, we critically evaluate various hand gesture recognition approaches for the tactile interface based on textiles. We propose a standardized framework for processing tactile signals obtained from the textile interface and a systematic data collection protocol that improves the robustness of data-driven recognition methods. We developed and designed a textile-based tactile sensing system to evaluate the hand-gesture recognition methodologies. 





\section{Background \& Related Work}
\label{sec:related_work}


The sense of touch in human-machine interaction refers to the ability of a machine or device to detect and respond to tactile stimuli, much like the human sense of touch \cite{kaboli2015humanoids, kaboli2016tactile}. This interaction involves the use of tactile sensors or interfaces that enable machines to perceive physical contact, pressure, texture, temperature, and other tactile sensations \cite{murali2022active}. In human-machine interaction, incorporating the sense of touch enhances the user experience by providing more natural and intuitive ways of communication and control \cite{witchuda2023reservoir, thongking2022implementation}. It allows users to interact with machines in a manner that mimics real-world tactile interactions, such as pressing buttons, gesturing, or manipulating objects. By integrating tactile feedback into human-machine interfaces, devices can provide users with sensory information that complements other forms of feedback, such as visual or auditory cues. This can improve usability, safety, and efficiency in various applications, including robotics, virtual reality, gaming, healthcare, and automotive systems. The sense of touch is essential for understanding and interacting with objects and other humans, allowing one to distinguish their physical characteristics. It plays a vital role in self-awareness, being able to differentiate the “me” from “not me.” The absence of the sense of touch would widen the gap between what is sensed (the "raw" data received by the human senses) and what is perceived (the brain`s interpretation of what the "raw" data is) \cite{NeuroInspiredElectronicSkinForRobots, AReviewofTwactileInformation}. 

Significant advancements have been made in the realm of intelligent systems to enhance human-robot interaction (HRI) and human-machine interaction (HMI), recognizing their pivotal role in social communication, especially non-verbal exchanges. A variety of devices, such as touchscreens, tactile interfaces, wearable devices, and sensors, have been leveraged to facilitate these interactions \cite{TheGrenobleSystemfortheSocialTouchChallengeatICMI2015touch-screen, AutomaticRecognitionofTouchGesturesintheCorpusofSocialTouchJung2017, DesignofaTherapeuticRoboticCompanionForRelationalAaffectiveTtouchStiehl2005DesignOA, ArtificialSkinAndTactileSensingForSociallyInteractiverobotsSILVERATAWIL2015230}. Such efforts underscore the importance of touch in conveying social cues and mood among humans \cite{TheCommunicationofEmotionViaTouchemotion}. A range of hardware technologies have been investigated to develop artificial touch systems. In recent decades, researchers have devoted considerable effort to developing these systems using diverse sensor technologies, including resistive, piezoresistive, capacitive, optical, piezoelectric, and acoustic sensors. These sensors can operate individually or in combination to capture various aspects of touch \cite{NeuroInspiredElectronicSkinForRobots}. Despite these strides, replicating the intricate touch-sensing capabilities of humans remains a formidable challenge.

Capacitive sensing has established itself as a dominant technology in this field, having gained widespread acclaim since its integration into the Apple iPhone in 2007 \cite{ProjectedCapacitiveTouchTechnology}. Consequently, mutual-capacitive touchscreens have become the predominant touch-sensing technology for mobile devices \cite{InvestigatingtheFeasibilityofFingerIdentificationonCapacitiveTouchscreensUsingDeepLearning}.

Investigations are being conducted into innovative methods for users to engage in varying manners and on diverse surfaces. For example, \cite{ArobustOnlineTouchPatternRecognitionforDynamicHuman-robotInteraction} evaluates the algorithmic effectiveness of touch interactions on robots, specifically utilizing the KAIST Motion Expressive Robot (KaMERo). Additionally, \cite{InterpretationoftheModalityOfTouchOnAnArtificialArmCoveredWithAnEIT-basedSensitiveSkin} employs a flexible and stretchable artificial skin based on the principles of electrical impedance tomography for robotic applications. Meanwhile, \cite{RecognizingHumanTouchingBehaviorsUsingaHapticInterfaceForaPet-robot} utilizes a haptic interface constructed from gridded pressure-sensitive conductive ink sheets for a pet-robot interaction.

Within touch gesture recognition, a multitude of methods have been explored, revolving around feature extraction or feature-engineering. Nearest-Neighbor (NN) classification has effectively facilitated gesture recognition, as indicated by multiple studies, \cite{Protractor:AFastandAccurateGestureRecognizer,SearchingandMiningTrillionsofTimeSeriesSubsequencesUnderDynamicTimeWarping,TheEffectofSamplingRateonthePerformanceofTemplate-basedGestureRecognizers,GesturesAsPointClouds:APRecognizerforUserInterfacePrototypes}. Other prominent methods, such as statistical classifiers, are also prevalent. For example, \cite{ThePowerofAutomaticFeatureSelection:RubineonSteroids} uses the Rubine Algorithm (a simple feature extractor and linear classifier \cite{rubine1991} designed specifically for Gesture Recognition) to assemble, categorize and selectively choose the best features from a comprehensive feature library for a specified problem. Some studies employ Hidden Markov Models (HMMs) for pattern recognition, such as \cite{HMM-basedEfficientSketchRecognition} or Parametric HMMs such as \cite{ParametricHiddenMarkovModelsforGestureRecognition}. 
However, techniques leveraging Hidden Markov Models possess several drawbacks, including the need for conditional independence of observations and the challenge of selecting appropriate hidden states within a generative model. With machine learning techniques, as shown in \cite{IconicandMulti-strokeGestureRecognition}, support vector machine (SVM), multi-layer perceptron (MLP), and discrete wavelet transform (DWT) have shown reasonable performances in classical touch interfaces such as touch screens. 

The recent focus of researchers has been on deep learning-based techniques, such as convolutional neural networks (CNNs) and recurrent neural networks (RNNs). The authors in \cite{ShallowConvolutionalNeuralNetworksforHumanActivityRecognitionUsingWearableSensors} presented a CNN-based approach to social gestures\cite{TouchingtheVoid10.1145/2663204.2663242}. Similarly, researchers in \cite{TouchModalityClassificationUsingRecurrentNeuralNetworks} examine the time series properties of RNNs to exploit the temporal characteristic of touch signals for tactile classification, with a specific emphasis on the analysis of long-term dependencies within the data. The reported results suggest that such recurrent networks have significant potential to be used for touch-based gesture recognition. 


To the best of our knowledge, a systematic study on gesture recognition methods has not been performed on emerging textile-based touch interfaces, which is addressed in this work.

\section{Methodologies}
In this study, we undertake an evaluation of different hand gesture recognition approaches for a large area tactile sensing interface constructed from conductive textiles. We explore both traditional feature engineering methods and contemporary deep learning techniques to address the challenge of gesture recognition. Leveraging insights from a comprehensive state-of-the-art review, we identify and analyze five prominent algorithms. Furthermore, we devised a standardized experimental protocol to assess the robustness of these methods. Here, `robustness' refers to the capability of the methods to maintain performance across diverse conditions, including variations in hand sizes, movement speeds, applied pressures, and interaction locations. This section provides a detailed exposition of our comparative analysis conducted on the developed textile-based touch interface.
\subsection{Hardware: Textile-based Tactile Interface}
\label{sec:Hardware}
The tactile interface utilized for this study is the \textbf{TexYZ} textile-based capacitive sensor \cite{TeXYZ}. Uniquely crafted through an embroidery process, this cutting-edge textile tactile surface leverages mutual-capacitance sensing to pinpoint finger interactions on the hand rest, delivering a spatial resolution of $9\times9$ taxels. The authors in \cite{TeXYZ}, opted to employ mutual-capacitive sensing as opposed to the more rudimentary method of self-capacitance providing: three-degree-of-freedom detection (2D location and pressure $xyz$), multi-touch detection and being less sensitive to EM noise. Furthermore, due to the natural flexibility provided by the textile sensor, it can be seamlessly integrated with any surface, endowing it with tactile properties and expanding its interactive capabilities. \cite{TeXYZ} utilizes a Cypress Programmable System-on-Chip (PSoC) \cite{cypress}. The PSoC board is a development board that hosts a PSoC micro-controller unit (MCU) and integrates configurable analog and digital peripheral functions, memory, and a microprocessor on a single chip. 
The UDP transmission approach was used to transmit data between the sensor and the software, over a local network with the OSC protocol \cite{osc} at a rate of 15Hz. 

\begin{figure}[t!]
    \centering
    \includegraphics[width=0.7\columnwidth]{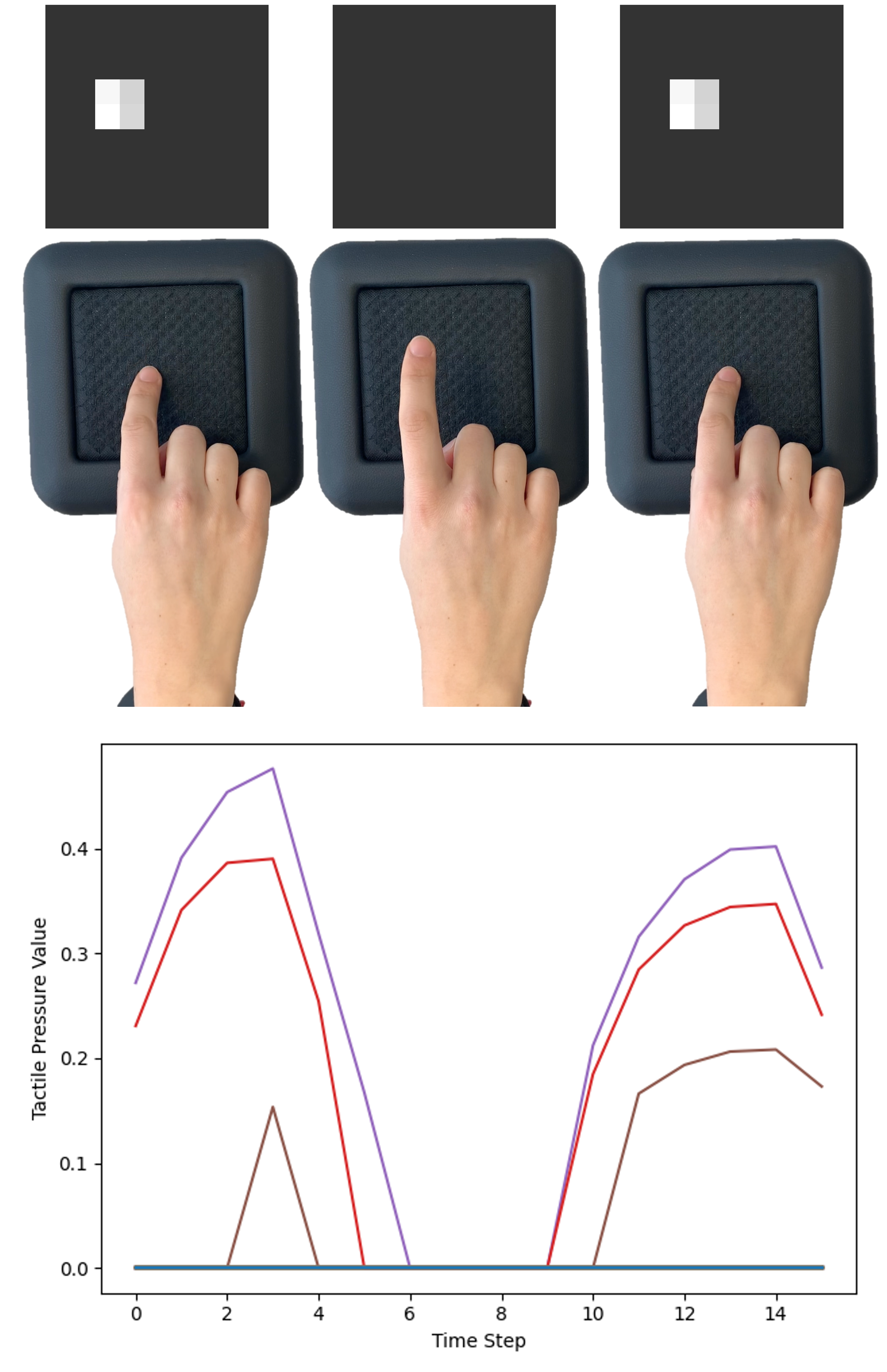}\label{fig:tacell1}
    \caption{Example of the gesture "double tap" being performed on the touch interface \cite{TeXYZ}.
    The bottom graph represents the evolution through time of the pressure values of each taxel (each one represented with a different color). Each matrix of the top graphs represents a snapshot of the pressure values on the sensor}
\end{figure}

\subsection{Hand Gesture Protocols}
\label{sc:Data Collection}

To effectively evaluate the various methods of hand gesture recognition, it is essential to clearly define a set of natural gestures that users would typically perform to interact with the touch interface. To achieve this, we established a repertoire of ten distinct gestures, each varying in complexity. Subsequently, data collection sessions were conducted with the participation of 34 individuals.  To test the robustness of the recognition approaches, data collection was carried out under three different inclinations of the tactile interface: parallel to the ground, tilted 30 degrees towards the participant, and tilted 60 degrees towards the participant. In addition, participants were instructed to perform each gesture at three different speeds: regular, fast, and slow. The duration of the recording of each gesture was adjusted to match the length of the gesture, ensuring that the perceived speed of the gesture was consistent and natural across participants. This led to gesture data having variable length.  

The gesture data collected from each participant was filtered (running average) and normalized. During these sessions, participants were instructed to perform the following touch gestures as part of the data collection process.
\begin{itemize}
    \item One Tap: a single tap on the surface.
    \item Double Tap: two quick taps on the surface.
    \item Swipe Down: dragging the finger toward the lower part of the sensor.
    \item Swipe Up: dragging the finger toward the upper part of the sensor.
    \item Swipe Right: dragging the finger toward the right side of the sensor.
    \item Swipe Left: dragging the finger toward the left side of the sensor.
    \item Circle Clock-wise: making a circular motion in a clockwise direction.
    \item Circle Counter Clock-wise: making a circular motion in a counter-clockwise direction.
    \item Swipe Up Two Fingers: dragging two fingers towards the upper part of the sensor.
    \item Swipe Down Two Fingers: dragging two fingers towards the lower part of the sensor.
\end{itemize}

\begin{figure}[h!]
    \subfloat[Single tap]{\includegraphics[width=0.09\textwidth]{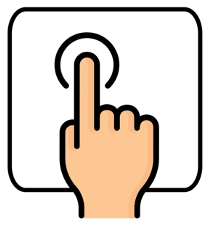}\label{fig:gesture-icon-0}}\hfill
    \subfloat[Double\\tap]{\includegraphics[width=0.09\textwidth]{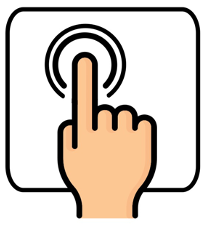}\label{fig:gesture-icon-1}}\hfill
    \subfloat[Swipe\\down with\\one finger]{\includegraphics[width=0.09\textwidth]{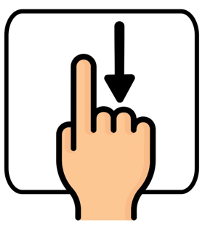}\label{fig:gesture-icon-2}}\hfill
    \subfloat[Swipe up\\with one\\finger]{\includegraphics[width=0.09\textwidth]{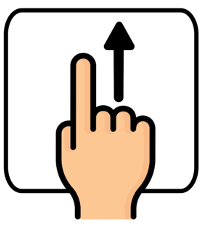}\label{fig:gesture-icon-3}}\hfill
    \subfloat[Swipe\\right with\\one finger]{\includegraphics[width=0.09\textwidth]{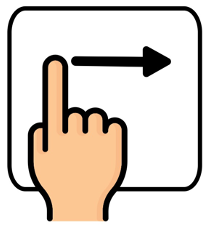}\label{fig:gesture-icon-4}} \\
    \subfloat[Swipe left\\with one\\finger]{\includegraphics[width=0.09\textwidth]{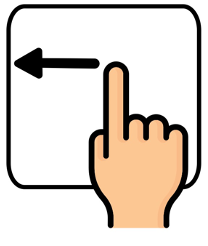}\label{fig:gesture-icon-5}}
    \hfill
    \subfloat[Circle\\clockwise]{\includegraphics[width=0.09\textwidth]{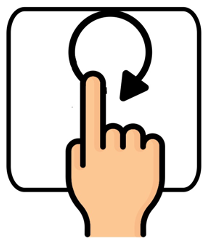}\label{fig:gesture-icon-6}}
    \hfill
    \subfloat[Circle\\counter-\\clockwise]{\includegraphics[width=0.09\textwidth]{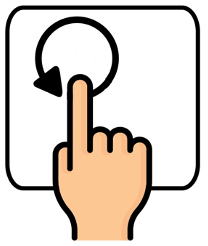}\label{fig:gesture-icon-7}}
    \hfill
    \subfloat[Swipe up\\with two\\fingers]{\includegraphics[width=0.09\textwidth]{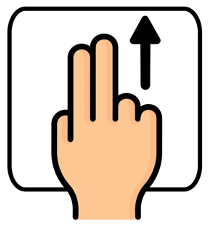}\label{fig:gesture-icon-8}}
    \hfill
    \subfloat[Swipe\\down with\\two fingers]{\includegraphics[width=0.09\textwidth]{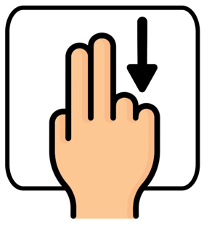}\label{fig:gesture-icon-9}}
    \caption{The 10 Hand Gesture movements associated with the respective class}
    \label{fig:hand gesture icons}
\end{figure}

\begin{figure}[h!]
    \subfloat[Single tap]{\includegraphics[width=0.09\textwidth]{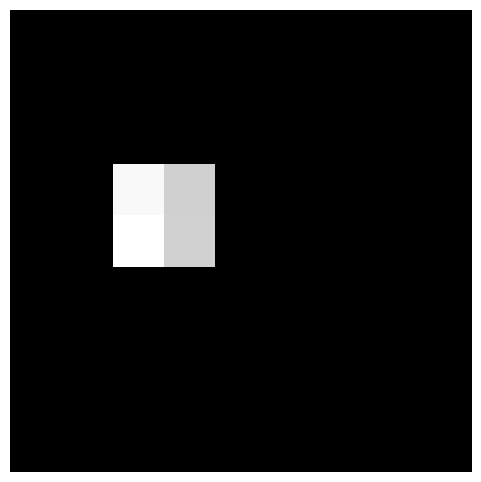}\label{fig:gesture0}}\hfill
    \subfloat[Double\\tap]{\includegraphics[width=0.09\textwidth]{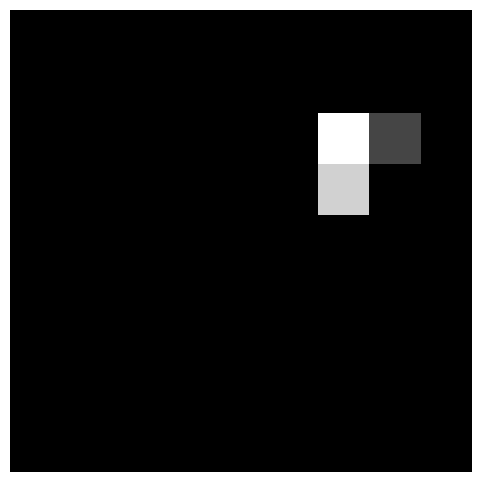}\label{fig:gesture1}}
    \hfill
    \subfloat[Swipe\\down with\\one finger]{\includegraphics[width=0.09\textwidth]{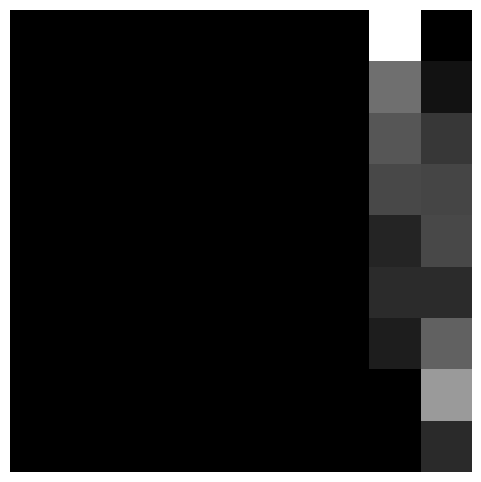}\label{fig:gesture2}}
    \hfill
    \subfloat[Swipe up\\with one\\finger]{\includegraphics[width=0.09\textwidth]{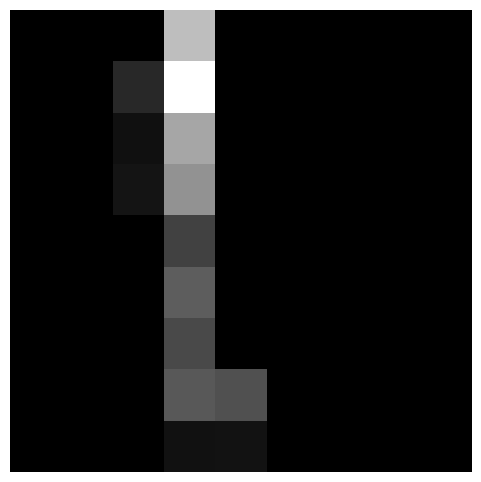}\label{fig:gesture3}}\hfill
    \subfloat[Swipe\\right with\\one finger]{\includegraphics[width=0.09\textwidth]{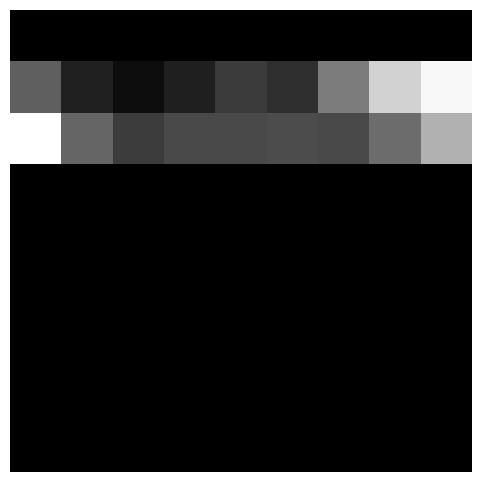}\label{fig:gesture4}}\hfill \\
    \subfloat[Swipe left\\with one\\finger]{\includegraphics[width=0.09\textwidth]{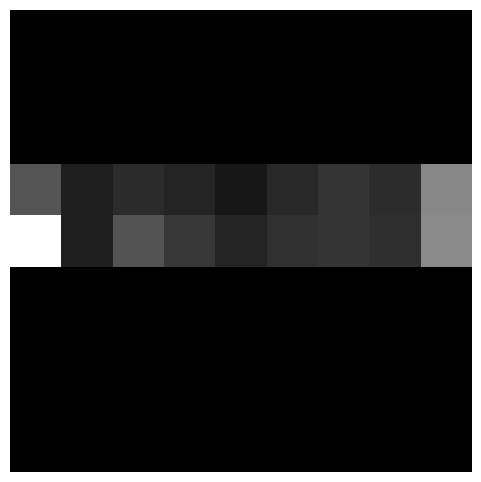}\label{fig:gesture5}}
    \hfill
    \subfloat[Circle\\clockwise]{\includegraphics[width=0.09\textwidth]{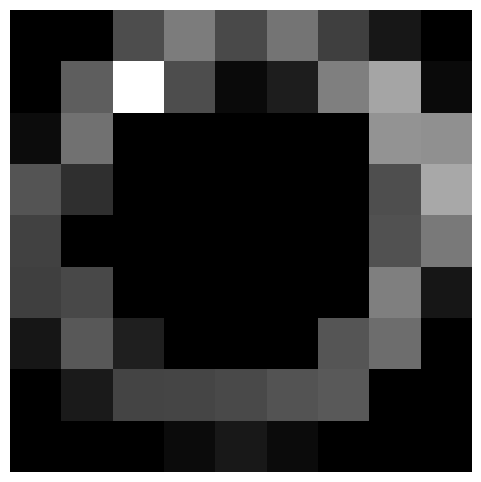}\label{fig:gesture6}}
    \hfill
    \subfloat[Circle\\counter-\\clockwise]{\includegraphics[width=0.09\textwidth]{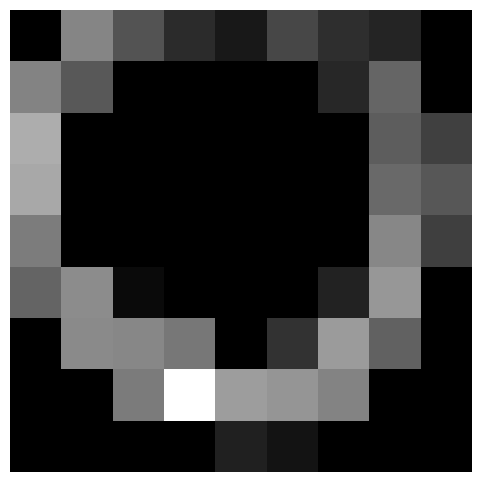}\label{fig:gesture7}}
    \hfill
    \subfloat[Swipe up\\with two\\fingers]{\includegraphics[width=0.09\textwidth]{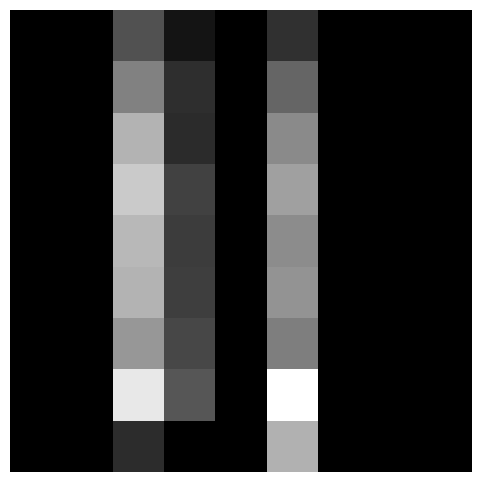}\label{fig:gesture8}}
    \hfill
    \subfloat[Swipe\\down with\\two fingers]{\includegraphics[width=0.09\textwidth]{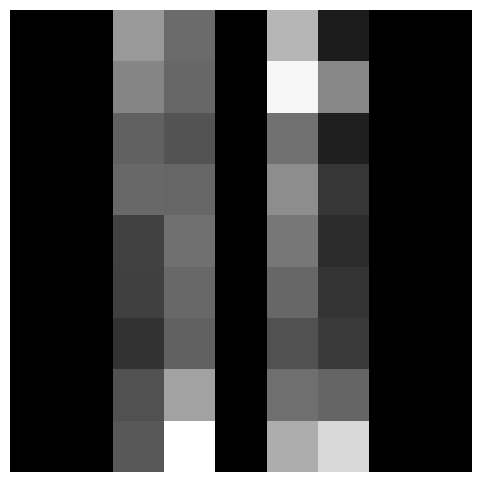}\label{fig:gesture9}}
    \caption{Hand Gestures Raw Signal from the \textbf{TexYZ}, summed over the whole time series length}
    \label{fig:raw signals}
\end{figure}

These ten gestures are depicted in Figure \ref{fig:hand gesture icons}, with their associated raw signals depicted in Figure \ref{fig:raw signals}.

We obtained two forms of temporal (time-series) gesture data: raw pressure data and trajectory (position of the finger over time) data. The pressure data consists of pressure values from each taxel in a given time frame, resulting in pressure values of $N\times N \times T$ (where N is the dimension of the tactile sensor interface, 9 in this specific case, and $T$ is the duration of the gesture). The trajectory data consisted of the $x$ and $y$ coordinates of up to three detected fingers. These coordinates are defined relative to an origin in the bottom left corner of the sensor array, with $x$ increasing to the right and $y$ increasing upward. The trajectory data are computed through processing performed on the PSoC. In total, we collected $3060$ time-series gesture data. 

\subsection{Data Augmentation}
\label{sec:data augm}
The methods selected for evaluation (feature-engineered as well as deep-learning), require a reasonable amount of gesture data for satisfactory performance. However, collecting large-scale user gesture data is time-consuming and resource-intensive. To address the limited amount of available gesture data, the following algorithm has been developed to augment the gesture data. It was imperative to execute data augmentation carefully to ensure that the gesture remained intact without any truncation. The primary objective was to achieve the shift-invariance (both vertically and horizontally) for the gestures. This would enable the model to recognize a gesture irrespective of its location on the tactile sensor. We also evaluated the gesture-recognition methods with and without the proposed augmentation of the collected gesture data. 

\begin{algorithm}
\caption{Dataset Augmentation by Shifting}
\begin{algorithmic}[1]
\REQUIRE{$\mathbf{D}$ (Original dataset), $\mathbf{L}$ (Corresponding labels)}
\ENSURE{$\mathbf{D_{\text{aug}}}$ (Augmented dataset), $\mathbf{L_{\text{aug}}}$ (Augmented labels)}
\STATE $\mathbf{D_{\text{aug}}} \leftarrow \emptyset$
\STATE $\mathbf{L_{\text{aug}}} \leftarrow \emptyset$
\FOR{each gesture $g$ in $\mathbf{D}$}
    \STATE Calculate bounding box of $g$ as $(x_{tl}, y_{tl}, x_{br}, y_{br})$  
    \STATE $shift_{\text{right}} \leftarrow 9 - x_{br} - 1$
    \STATE $shift_{\text{left}} \leftarrow x_{tl}$
    \STATE $shift_{\text{up}} \leftarrow y_{tl}$
    \STATE $shift_{\text{down}} \leftarrow 9 - y_{br} - 1$
    \STATE Append $g$ and its shifted versions by $shift_{\text{right}}, shift_{\text{left}}, shift_{\text{up}}, shift_{\text{down}}$ to $\mathbf{D_{\text{aug}}}$
    \STATE Append label of $g$ for each augmented version to $\mathbf{L_{\text{aug}}}$
\ENDFOR
\RETURN $\mathbf{D_{\text{aug}}}, \mathbf{L_{\text{aug}}}$
\end{algorithmic}
\end{algorithm}

\renewcommand\thesubfigure{} 



 \subsection{Hand Gesture Recognition Methods}
\label{ch:methodology}

\begin{figure*}[tbh!]
    \centering
    \includegraphics[width=0.95\textwidth]{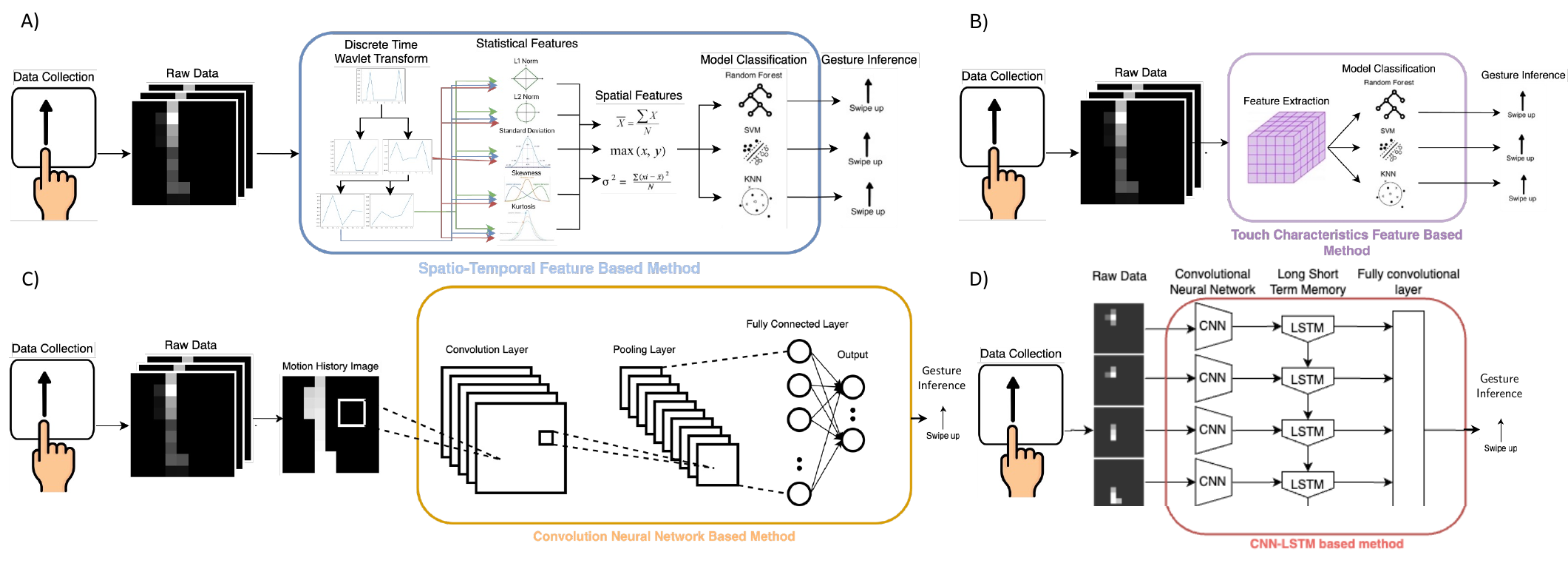}
    \caption{Illustration of the hand gesture recognition methods evaluated in this work.}
    \label{fig:comparision_method}
\end{figure*}

In this section, we present a brief overview of the various hand gesture recognition approaches utilized for comparison in this study. 

\subsubsection{Spatio-temporal Features Approach}
\label{Spatio-Temporal Feature Approach}
In this method, we extract spatial and temporal features from the raw pressure data obtained from each gesture, drawing inspiration from the approach described in \cite{TouchGestureRecognitionUsingSpatiotemporalFusionFeatures9459742}. Leveraging the time-dependent nature of the data, we employ Discrete Wavelet Transforms (DWT) for feature extraction. DWT is applied to the raw signal obtained from each taxel across the entire time series, yielding a set of coefficients that capture various signal characteristics. 

For each wavelet coefficient, we compute statistical features such as the L1 norm, L2 norm, skewness, kurtosis, and standard deviation.  We also introduced three spatial features: Mean, Max, and Variance of all the taxels, to obtain correlation information among the taxels during each gesture. An illustrative overview of the approach is presented in  Fig.\ref{fig:comparision_method}(A). The computed features are then utilised within different classification algorithms, which are presented in the following section.

\subsubsection{Touch Pattern Recognition Approach}
\label{Touch Pattern Recognition Approach}
In this approach, we derive descriptive features (illustrated in Fig.\ref{fig:comparision_method}(B)) from three key parameters of touch gestures: pressure intensity, approximate contact surface area, and duration of gestures. Drawing insights from previous research \cite{AutomaticRecognitionofTouchGesturesintheCorpusofSocialTouchJung2017, TowardsSocialTouchIntelligence10.1145/2663204.2666281, TouchingtheVoid10.1145/2663204.2663242}, the features include mainly mean pressure, which aggregates pressure readings across all taxels over time, and maximum pressure, which captures the highest recorded value across all taxels during the gesture. In addition, pressure variability is computed as the average difference in absolute values between sequential frames across all channels, taxels to capture fluctuations in the pressure throughout the gesture. 

In addition to these temporal metrics, spatial features of the pressure distribution are also computed as row-wise mean pressure, which is calculated to depict the pressure distribution along the length of the sensor array, while column-wise mean pressure represents the distribution along its width. Furthermore, the feature that captures the contact area is also included, which is the maximum and average contact area per frame during the gesture. This helps to distinguish between different types of touch, such as two-finger vs. single-finger gestures.

The feature extraction approach described above \ref{Spatio-Temporal Feature Approach} and \ref{Touch Pattern Recognition Approach} was used for classification with different classifiers: K-Nearest Neighbors (KNN), Random Forest (RF), and Support Vector Machine (SVM). The hyperparameters of each were optimized in the train/validation set using cross-validation leave-one-subject-out (5 times). The results of gesture recognition were evaluated and the best performing hyperparameters are presented in Table \ref{table:parameters}.

\subsubsection{Deep Learning (DL) Approach- CNN}
Deep learning (DL) has surged in popularity, often outperforming traditional methods that rely on feature engineering. Among DL methods, Convolutional Neural Networks (CNNs) have emerged as particularly powerful, especially for array-like data such as images. However, CNNs are inherently designed for static images, which poses a challenge when dealing with dynamic time-series data, such as gestures. To overcome this limitation, we utilize Motion History Images (MHIs) derived from the raw data. An MHI is a grayscale image in which recently activated taxels appear brighter, effectively capturing the motion dynamics of gestures \cite{UsingMotionHistoryImageswith3DConvolutionalNetworksinIsolatedSignLanguageRecognition}. This approach enables the network to effectively learn and classify dynamic gesture data by leveraging the temporal information encoded in the MHIs. 

Our selected CNN architecture, as depicted in Fig.\ref{fig:comparision_method}(C), comprises four convolutional layers with Leaky ReLU activation functions, followed by two fully connected layers. The final layer comprises a 10-logit output representing the set of gesture labels.
\subsubsection{DL Approach- Recurrent Neural Network}
\label{subs:RNN}



RNNs are a widely recognized deep learning architecture that is well suited to handle time series and sequential data, with applications that include gesture recognition, as demonstrated in \cite{TouchModalityClassificationUsingRecurrentNeuralNetworks, LSTMFullyConvolutionalNetworksforTimeSeriesClassification}. However, classical RNNs often encounter challenges such as vanishing gradients, limiting their ability to capture long-range dependencies effectively. To address this, LSTM (Long Short-Term Memory) networks have been introduced, designed to mitigate the limitations of traditional RNNs. In this study, we employ a standard LSTM implementation \cite{LSTMFullyConvolutionalNetworksforTimeSeriesClassification}. Each $9 \times 9$ gesture data matrix in each time frame was flattened into a vector of dimensions $81$, treating each taxel as an individual feature input to the LSTM network. To avoid overfitting, we set the number of hidden units in the network to 32. Additionally, given that gestures in our dataset vary in length based on their duration, it was essential to standardize the sequence length. We achieve this by padding the gesture time series to ensure a uniform sequence length. A dense fully connected layer with 10-logit output was added after the LSTM layer to facilitate gesture classification. This sequential processing allows the network to recognize the dynamics of gestures, such as speed and direction of motion.  

\subsubsection{DL Approach- Convolutional Neural Network + Long Short Term Memory}
\label{sec: CNN LSTM}
We observed that the CNN model struggles to discriminate between gestures of `\textit{one tap}' and `\textit{double tap}' because the MHI images for those two gestures are extremely similar. Furthermore, the LSTM model with flat input does not take advantage of the spatial characteristics of gestures. Therefore, to combine the strength of both of these architectures, we evaluated the integration of Convolutional Neural Networks (CNNs) with Long Short-Term Memory (LSTM) \cite{LSTMFullyConvolutionalNetworksforTimeSeriesClassification}. The CNN-LSTM architecture is particularly suitable for the gesture recognition problem, where the input data is both spatially and temporally rich, making it ideal for interpreting tactile sensor data with spatio-temporal dynamics. 

In our CNN-LSTM network (illustrated in Fig.\ref{fig:comparision_method}(D)), the initial layers are convolutional to extract spatial features, including patterns that are indicative of the type of gesture performed. In contrast to the previous CNN approach, the input is not Motion History Images (MHIs) but the $9 \times 9$ arrays produced from every time frame. The CNN is composed of two convolutional layers, with ReLu as an activation function and a Max-Pooling layer. After the convolutional layers, the output feature maps are flattened and fed into the LSTM layers. The LSTM layers are designed with 32 hidden units to handle sequences considering the temporal evolution of gestures. This enables the network to learn not only from the static patterns within each frame but also from the transition of these patterns over time. This combination of spatial feature extraction via CNN and temporal sequence processing via LSTM provides the CNN-LSTM model to effectively classify/recognize complex hand gestures and outperforms other DL approaches, as presented in Sec \ref{sec:offline_evaluation}. The output of the LSTM is then fed to a fully connected layer of size 10 corresponding to the gestures.  


\section{Results and Discussion}
\label{sec:results}
In this section, we present the results of the evaluation of the hand-gesture recognition methods described above for the textile-based touch interface. We performed both an offline evaluation on the collected data set and an online evaluation on new participants to evaluate not only the recognition performance but also the real-life use case. We evaluated the gesture recognition capability in terms of classification accuracy, i.e. the ratio of correct gesture label prediction to total gestures performed.  


\subsection{Offline Evaluation}
\label{sec:offline_evaluation}
\begin{figure*}[t!]
\centering
    \includegraphics[width=\textwidth]{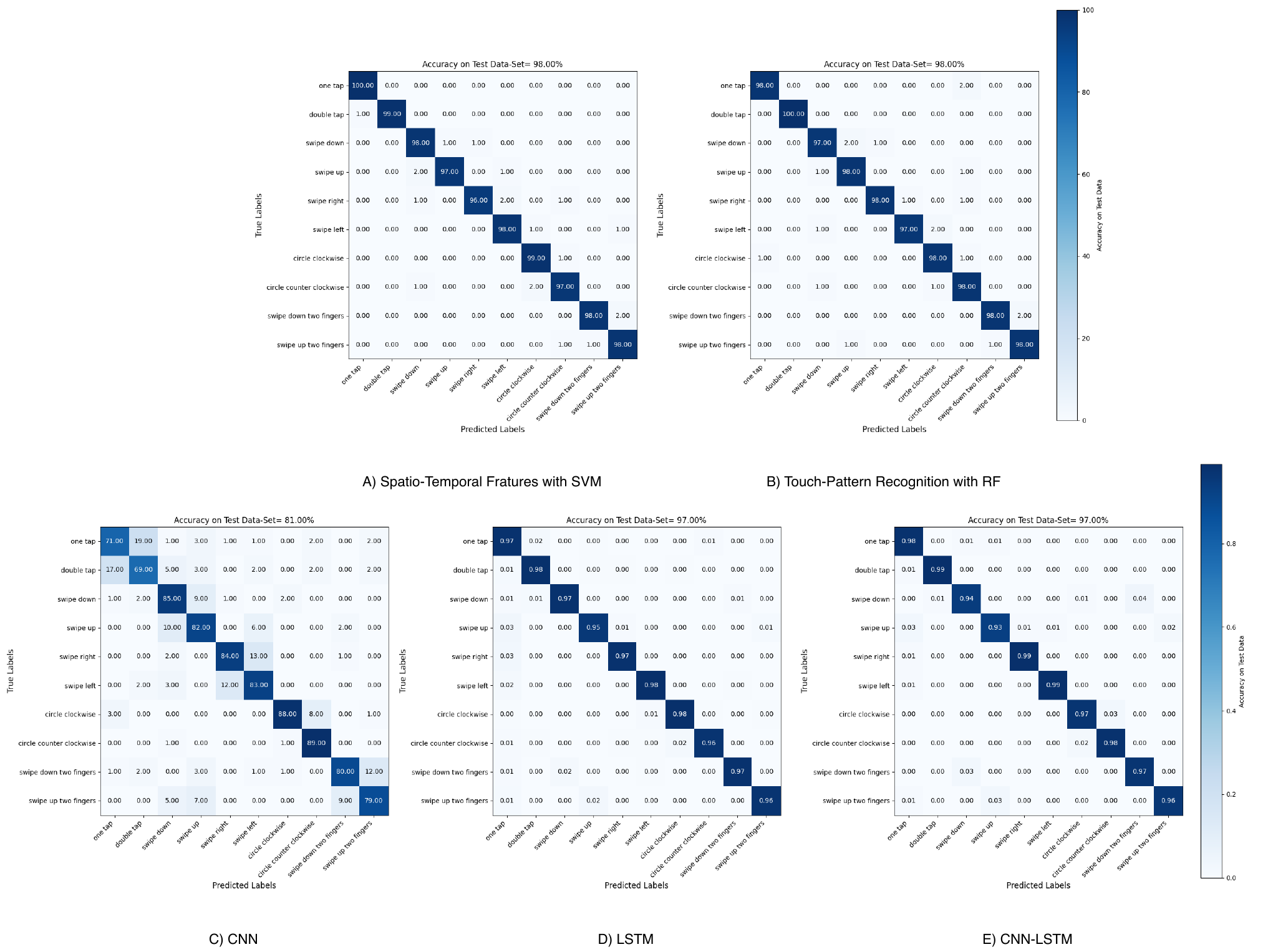}
    \caption{Confusion matrices of the best performing configuration for each method in the offline evaluation}
    \label{fig:conf_mtx}
\end{figure*}

For offline evaluation, we utilised 85\% of the collected hand gestures for training and 15\% for testing. We report the classification accuracy of feature engineering-based methods - \textit{ spatial-temporal features approach} and \textit{touch pattern recognition approach} in Table \ref{table:fe_results}. We critically evaluated each feature set with different classifiers KNN, RF, SVM and also the effect of data augmentation on classification performance.
In addition, we also present the confusion matrix of the best-performing classifier approach in Fig.\ref{fig:conf_mtx}. We can observe that without the application of \textit{data augmentation}, the classification performance varies between $0.58$ and $0.62$ for the \textit{ spatial-temporal} method and between $0.56$ and $0.6$ for the \textit{touch pattern recognition} method. However, with \textit{data augmentation}, both approaches achieve accuracy up to $98\%$. Analyzing the confusion matrices reveals a consistent trend across feature-engineering methods: difficulty in distinguishing between clockwise and counterclockwise circles, as well as swipe gestures in various directions. In the \textit{spatio-temporal} approach, this difficulty likely originates from the feature extraction process. The statistical features derived from Discrete Wavelet Transform (DWT) coefficients evaluate values across the entire signal. Given that swiping gestures in opposite directions can exhibit similar total signal magnitudes, regardless of the frame order in the original raw signal, distinguishing between them becomes challenging. Similarly, the \textit{touch pattern recognition} method faces difficulties with similar gesture designs. Analogous gestures, such as swiping up and down or left and right, have nearly identical duration and coverage in the touch area.

Table \ref{table:dl_results} presents the classification accuracy of the deep learning-based approach, along with the confusion matrices of the best performing versions in Fig.\ref{fig:conf_mtx}. We can observe that the CNN approach does not perform satisfactorily using the MHIs technique, having a high level of confusion between gestures `\textit{single tap}' and `\textit{double tap}'. Both LSTM and CNN-LSTM demonstrated similar performance, with CNN-LSTM having a slight edge when no data augmentation was utilised signifying efficient recognition performance. 

\begin{table}[t!]
    \centering
    \begin{tabular}{@{}lccc@{}}
        \toprule
        Method                              & Data Augmentation & Classifier & Accuracy \\
        \midrule
        Spatio-temporal Features            & \xmark            & KNN        & 0.62     \\
        Spatio-temporal Features            & \xmark            & RF         & 0.58     \\
        Spatio-temporal Features            & \xmark            & SVM        & 0.58     \\
        Spatio-temporal Features            & \cmark            & KNN        & 0.79     \\
        Spatio-temporal Features            & \cmark            & RF         & 0.96     \\
        Spatio-temporal Features            & \cmark            & SVM        & 0.98     \\
        Touch Pattern Recognition           & \xmark            & KNN        & 0.56     \\
        Touch Pattern Recognition           & \xmark            & RF         & 0.60     \\
        Touch Pattern Recognition           & \xmark            & SVM        & 0.56     \\
        Touch Pattern Recognition           & \cmark            & KNN        & 0.98     \\
        Touch Pattern Recognition           & \cmark            & RF         & 0.98     \\
        Touch Pattern Recognition           & \cmark            & SVM        & 0.97     \\
        \bottomrule
    \end{tabular}
    \caption{Table reporting accuracy results for Feature-Engineering methods with offline evaluation}
    \label{table:fe_results}
\end{table}

\begin{table}[t!]
    \centering
    \begin{tabular}{@{}lcc@{}}
        \toprule
        Method                                  & Data Augmentation  & Accuracy \\
        \midrule
        CNN                 & \xmark               & 0.77     \\
        CNN                 & \cmark             & 0.81     \\
        RNN with LSTM       & \xmark             & 0.82     \\
        RNN with LSTM       & \cmark             & 0.97     \\
        CNN and LSTM        & \xmark             & 0.89     \\
        CNN and LSTM        & \cmark             & 0.97     \\
        \bottomrule
    \end{tabular}
    \caption{Table reporting accuracy results obtained for Deep Learning methods with offline evaluation}
    \label{table:dl_results}
\end{table}

\begin{table}[t!]
    \centering
    \begin{tabular}{@{}lcc@{}}
        \toprule
        Method                               & Data  &  Accuracy \\
                                            &  Augmentation   &  \\

        \midrule
        Touch Pattern         & \cmark            & 0.71     \\
        Recognition RF        &  &       \\

        RNN with LSTM          & \cmark             & 0.89     \\
        CNN and LSTM           & \cmark             & 0.93     \\
        \bottomrule
    \end{tabular}
    \caption{Table reporting accuracy results obtained for the selected model with online evaluation}
    \label{table:results_online}
\end{table}

\begin{table}[t!]
    \centering
    \begin{tabular}{@{}lcccl@{}}
        \toprule
        Method                        & Data  & Classifier & Parameter         \\
                                    &  Aug. &  &                                   \\

        \midrule
        Spatio-temporal Features      & \xmark            & KNN        & K = 7                                       \\
        Spatio-temporal Features      & \xmark            & RF         & NoE = 200, \\
        & & & max depth = 9   \\
        Spatio-temporal Features      & \xmark            & SVM        & C = $2^7$, $\gamma = 2^{-11}$               \\
        Spatio-temporal Features      & \cmark            & KNN        & K = 7                                       \\
        Spatio-temporal Features      & \cmark            & RF         & NoE = 200,\\ &&&max depth = 9   \\
        Spatio-temporal Features      & \cmark            & SVM        & C = $2^9$, $\gamma = 2^{-13}$               \\
        Touch Pattern Recognition     & \xmark            & KNN        & K = 4                    \\
        Touch Pattern Recognition     & \xmark            & RF         & NoE = 200, \\&&&max depth = 9   \\
        Touch Pattern Recognition     & \xmark            & SVM        & C = $2^{13}$, $\gamma = 2^{-13}$            \\
        Touch Pattern Recognition     & \cmark            & KNN        & K = 1                                       \\
        Touch Pattern Recognition     & \cmark            & RF         & NoE = 200,\\ &&&max depth = 9   \\
        Touch Pattern Recognition     & \cmark            & SVM        & C = $2^9$, $\gamma = 2^{-7}$                \\
        \bottomrule
    \end{tabular}
    \caption{Table reporting tuned hyperparameters for the different classifiers for feature-engineering approach, where NoE stands for Number of Estimators}
    \label{table:parameters}
\end{table}

\subsection{Online evaluation}
\label{sec: Online Evaluation}

In section \ref{sec:offline_evaluation}, we observed satisfactory classification accuracies, exceeding 97\% for most methods. However, these results diverged when tested in real time. In the offline evaluation, the models were trained using processed and filtered data that had limited noise or inherent fluctuations while performing gestures. Additionally, the online evaluation ensured that the gesture data obtained was from out-of-distribution of the training set, ensuring a robust evaluation. A diverse and representative sample of participants was considered for online testing, including varying hand sizes and demographic characteristics. We present the online recognition performance of the best performing methods (Touch Pattern Recognition with RF classifier, LSTM, and CNN-LSTM) in Table. \ref{table:results_online}. 
Although the \textit{spatio-temporal features} method managed to achieve a 98\% accuracy rate in offline tests, its efficiency drastically decreased during the initial online tests, making it unsuitable for online evaluation. Moreover, this method required all the time series data in the dataset to have identical length, contrasting other approaches that were invariant to gesture length/duration. In our online tests, we observed that the accuracy for the \textit{touch pattern recognition} dropped to 71\%  and 89\% for the LSTM model and 93\% for the CNN-LSTM model. One limitation of deep learning-based approaches is that the dataset is inherently tied to the specific hardware on which it was collected. Consequently, the introduction of new hardware requires the recollection of data and the retraining of the neural network or the application of domain adaptation techniques. Furthermore, we observed that subsequent uses of the deep learning method resulted in deteriorating performance which can be attributed to the data distribution shift.

\section{Conclusion}
In this paper, we present a comprehensive assessment of hand gesture recognition that has been validated on the in-house developed textile-based touch interface. We investigated traditional feature engineering methods and contemporary deep learning techniques to address the challenge of touch gesture recognition. Our findings indicate that the deep learning-based CNN-LSTM approach can recognize diverse sets of gestures consistently and robustly online, outperforming feature-based methods.
CNN-LSTM's method captures both spatial and temporal dependencies, a feature particularly advantageous for recognizing gestures that may involve similar movements, but differ in execution speed or sequence. Such nuances might otherwise be indistinguishable when relying solely on other deep learning- or feature-engineering-based approaches. In the future, our goal is to explore more realistic and extensive human gesture scenarios, examine one-shot learning methods for identifying completely novel gestures, and improve our inference capabilities to better accommodate real-time applications through long-term testing. In conclusion, our findings will drive advances in hardware and software for gesture recognition within touch interfaces and human-machine interactions.

\section{Acknowledgment}
We would like to thank Prajval Kumar Murali for his constructive suggestions and feedback.

\bibliography{references}
\bibliographystyle{IEEEtran} 

\end{document}